# TOWARDS A GENERIC SOFTWARE ARCHITECTURE FOR IOT SYSTEMS

Yousef Abuseta

Department of Internet Technologies, IT Faculty, University of Tripoli, Tripoli, Libya

## ABSTRACT

*The complexity of IoT, owing to the inherent distributed and dynamic nature of such systems, brings more challenges to the software development process. A vast number of devices with different communication protocols and data formats is involved and needs to be connected and exchange data with each other in a seamless manner. Traditional software architectures fall short of addressing the requirements of IoT systems and, therefore, a new approach to software architecture is required. This paper presents an attempt to lay out the foundation for a quality attribute driven software architecture for the development of IoT systems. This architecture accommodates the appropriate architectural styles and design patterns necessary for the development of a robust IoT system. These include edge computing, microservices and event driven architectures. The proposed architecture treats IoT systems as autonomic systems which require a closed control loop to regulate and orchestrate the operational aspect of the IoT system.*

## 1. INTRODUCTION

The Internet of Things (IoT) is the natural evolution of the Internet revolution that serves as a global architecture interconnecting machines and humans. The IoT paradigm revolves around designing software systems out of uniquely addressable things provided by sensors, actuators and processing capabilities to communicate and exchange data to accomplish a specific task [1]. This kind of connectivity enables a large number of applications in different domains such as healthcare, transportation, manufacturing, smart homes, etc. [2]. IoT systems are composed of heterogenous devices and applications in addition to communication networks for connecting them. It is estimated that by 2025, there will be over 75 billion connected devices worldwide [3], leading to more interactive and smarter environments.

Despite being widely adopted and highly promising, IoT systems still pose crucial challenges owing to the integrated heterogeneous software and hardware resources, communication protocols as well as data from various resources. A more challenging issue is the uncertainty that stems from the dynamic nature of the environment in which the IoT applications operate. Such a dynamicity takes the form of important events which include a device joining/leaving, introducing a new service or a change in system requirements.

These challenges may be addressed by a well-designed software architecture that accommodates the necessary components and appropriate design decisions. Therefore, designing an architecture that exhibits the qualities of supporting the development of a scalable, reliable, maintainable and extensible system is of paramount importance to successful IoT systems.

                                                                                    

A great deal of experiences of software development for IoT systems have been reported in literature. However, most proposed architectures for these systems have not emphasized enough the importance of the rationale for the choice of software components and how they relate to each other with regard to devising a robust IoT software architecture.

In this paper, we attempt to design a generic software architecture that supports the realization of the necessary quality attributes that are crucial for developing resilient IoT applications. This architecture consists of a set of appropriate structures, the relations between them as well as any constraints that control these relations. Also, the architectural styles, design patterns employed for achieving such an architecture are proposed here. The rest of this paper is organized as follows. Section 2 reviews related topics and concepts to the work conducted. In this paper Section 3 discusses the software architecture for IoT systems its characteristics and the important architectural and design patterns. Section 4 presents the proposed architecture. Section 5 discusses some related works. The paper is concluded in section 6 with some suggestions for future work.

## 2. BACKGROUND

### 2.1. A High Level IoT Layered Architecture

This section presents the IoT architecture layers along with the building blocks that constitute almost any IoT system. The IoT is built around the idea of connecting physical devices that are capable of exchanging information and can be monitored and controlled with an external system, usually an IoT platform. One of the main ideas of the IoT systems is the interaction between relevant users and physical devices, where sensors provide information to the user about the environmental conditions, and users can perform desired actions via actuators over other physical devices based on that information. Another key point of IoT systems is the possibility of the development of smart applications for specific domains. Examples of tasks such applications may carry out include data analytics, prediction, monitoring and user notification [4].

It is worth mentioning that there is no single agreed upon IoT architecture. Nevertheless, the key building blocks of any IoT system are almost the same which include smart things, networks and gateways, IoT middleware and applications.

These building blocks constitute the backbone of IoT systems on which an effective layered architecture can be developed. Such layers are described below and shown in Figure 1.

**The Sensing layer.** This layer represents the IoT system's point of contact with the physical world. Its main task is to translate physical information into data streams, or vice versa. The primary component of this layer is the physical device or thing. Physical devices include personal objects such as smart watches, and mobile phones. They also include objects in our environment (e.g. home, vehicle, etc.) and industry (e.g. machines, robots, motors). Devices are equipped with sensors and actuators. Sensors transmit the captured contextual environmental information using electrical signals to devices, to which they are attached.

This connection can be established either by wire or wirelessly [5]. Sensors may come in various forms such as motion sensors, location sensors, and temperature sensors or embedded sensors in industrial machines to gauge the machine's state [6]. An Actuator, on the other hand, is a hardware component which is used to apply necessary actions to the physical environment. It receives commands from its connected device and translates the received signals into some kind of physical actions [5].

**The Networking layer.** The main objective of this layer is to seamlessly connect IoT devices in the sensing layer with upper layers. It often contains a number of local gateways that can establish a connection with IoT devices using several network technologies such as Wi-Fi, Bluetooth, Zigbee and cellular networks. Data received from IoT devices is then converted from the device's protocol and forwarded to the Internet. An IoT gateway works as a bridge between the sensing layer and the Internet ensuring that data collected by IoT devices find its way to the Internet infrastructure. In addition to forwarding, gateways may also be tasked with performing some important operations, such as data aggregation and protocol translation [6].

**The Middleware layer.** The middleware layer is one of the primary components of the IoT platform as it serves as an integration layer for heterogeneous IoT devices and ICT systems in the platform. It facilitates application development by providing a uniform way of interaction with different hardware, protocols, and data formats. With a middleware layer in place, developers are relieved from the underlying hardware or network protocols and thus they concentrate fully on the application business logic. The main tasks of the middleware include: 1) Receiving data from the connected devices, 2) Processing the received data, 3) Providing the received data to connected Applications, and 4) Controlling devices [5]. Typically, an IoT Middleware can be accessed using APIs, for example, HTTP-based REST APIs.

**The Application layer.** An IoT application is a software component that carries out some tasks by reading data collected by IoT devices as well as and controlling these devices. IoT applications are used by end users to achieve data insights, control IoT devices and in general interact with the physical world using computers and smartphones. Applications can be built on top of the IoT platform or integrate with it through APIs. An IoT application can vary from a simple analytics dashboard where all IoT events are shown, to a complex application that processes data and extract information from IoT events and combine it with other information sources [6].

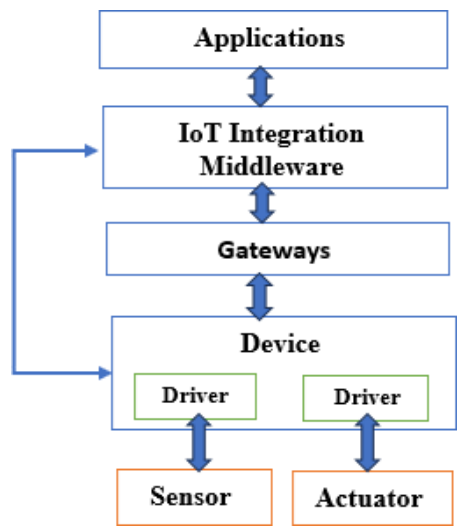

Figure 1. A high level IoT layered Architecture [7]

## 2.2. IoT Messaging Protocols

According to the IoT architecture depicted in Figure 1, multiple layers of communication can be recognized. In the first layer, communication occurs from IoT device to gateway, using protocols and radio technologies like Zigbee, Bluetooth, 802.15.4 and Wi-Fi. In the second layer, communication takes place from gateway to middleware. These include HTTP (REST), MQTT, AMQP, XMPP and CoAP. Event-based protocols such as MQTT are preferred, to prevent unnecessary costly polling. REST and MQTT are the most used protocols [8]. And finally, communication is conducted from middleware to applications, often using REST APIs [9].

The following are the most common protocols that are of great relevance to IoT systems. The discussion is based on a survey presented in [4].

- The Hypertext Transfer Protocol (HTTP): it is the primary the communications protocol forweb based applications. It is a generic stateless protocol which adopts the client–server model. It can be employed for many different tasks in addition to its use in the Web. In the IoT paradigm, HTTP can be used to exchange data with a publish or polling model.
- The Constrained Application Protocol (CoAP): it is a special protocol that is primely used in constrained environments (e.g., low-power devices). It is specifically designed for machine to machine (M2M) applications such as smart energy and building automation. Similar to HTTP, CoAP adopts the client–server model and it can be easily configured to communicate with HTTP endpoints.
- Message Queuing Telemetry Transport (MQTT): it is a lightweight publisher-subscriber protocol. MQTT is ideal for M2M and IoT contexts where a small code footprint is required and network bandwidth is at a premium.

## 3. SOFTWARE ARCHITECTURE FOR IOT SYSTEMS

### 3.1. Quality Attribute Driven Architecture

In [10], the software architecture is defined as: "*The software architecture of a system is the set of structures needed to reason about the system. These structures comprise software elements, relations among them, and properties of both*". Another definition in [11], defines software architecture as follows: "*software architecture reflects fundamental organization of a software system, its components and their relationships, as well as relations of a software system with its environment*". The key point in designing a resilient software architecture is to achieve loose coupling by reducing dependencies between involved structures. This simplifies software evolution as changes are often local and do not propagate through the architecture which highly contributes to the reduction of the cost and complexity of software maintenance [11]. In modern software systems, the software architecture is driven by quality attributes. The quality attributes are often referred to interchangeably as the non-functional requirements of the system under consideration. Such requirements are concerned with how well the system performs as opposed to the functional requirements which are related to the core functions the system in question must provide. In [12] the following definition of quality attributes was introduced: "*Quality attributes are properties of a software system and a subset of its non-functional requirements. Like other requirements, they should be measurable and testable. Software quality attributes are benchmarks that describe the software system's quality and measure the fitness of the system*". In another definition by [10], the quality attribute is defined as: "*A quality attribute (QA) is a measurable or testable property of a system that is used to indicate how well the system satisfies the needs of its stakeholders beyond the basic function of the system*". Examples of quality attributes include scalability, portability, security, availability and performance.

The quality attributes serve as a support tool for the evaluation of proposed software architecture. Various quality attributes are considered during the software design process. The satisfaction level of quality attributes impacts the overall performance of the system in consideration. [11].

Authors of [12, 13] stressed the importance of considering quality attributes in the early phase of eliciting software requirements. However, they stated that their testing and evaluation should be carried out during the whole system life cycle.

Quality attributes can be classified into two categories: the first category is concerned with those attributes that describe the system in question at runtime such as availability and performance while the second includes those that describe some property of the development of the system, such as modifiability, testability, or deployability [10].

A description of some important quality attributes is presented below.

**Interoperability.** This quality attribute can be defined as the ability of different technologies, software components or applications to seamlessly communicate with each other and interpret properly the exchanged data. In IoT, interoperability represents a serious issue and real challenge due to the heterogeneity of IoT systems and the lack of reference standards. [14].

**Performance.** It is about time and the software system's ability to meet timing requirements. In the context of IoT, this quality attribute is crucial as most IoT applications need to meet some desirable response time and throughput.

**Modifiability**. It is an attribute that enables a system to introduce a change. Changes may be required to: 1) introduce new features or modify existing ones, 2) fix defects, increase security and improve performance, 3) adopt new technology or protocols and 4) integrate different systems that were never designed to work together. [10].

**Energy Efficiency**. It is about managing energy consumption efficiently. This attribute is highly important as the majority of IoT devices, such as mobile phones and tablets, are energy constrained. Also, cloud providers are increasingly concerned with the energy efficiency of their server farms.

**Scalability**. It is about accommodating more of something Scalability is often needed to improve the system performance. Two kinds of scalability can be defined: horizontal and vertical. The horizontal scalability is concerned with accommodating more resources to logical entities, such as adding a new server to a server farm, whereas the vertical scalability refers to adding more resources to a physical entity, such as adding more memory to a computer.

Based on literature review conducted in [11], the following quality attributes were identified as the most important for IoT architecture: scalability, security, interoperability, and performance.

## 3.2. Architectural Styles for IoT Systems

The following discussion revolves around the realization of the above discussed quality attributes using some specific architectural styles and design patterns.

### 3.2.1. Edge computing

Over the last decade, the trend has been to use the centralized cloud computing infrastructure to provide data processing and storage in large data centres. This is due to its elasticity and cost-effectiveness. However, IoT systems pose significant challenges as many of these systems have devices (things) at the network edge that generate a tremendous amount of data that needs to be stored and processed using heavy resources. Current cloud computing is not efficient to process very large data in a timely manner to satisfy the end user and system requirements. Any failure by the cloud to comply with the timing requirement will likely have a negative impact on the quality of service (QoS) and affect the IoT applications and network performance in general. [15].

The edge computing has been proposed to address this issue. It is designed to bridge the gap between data generation and data processing. By processing data close to the source of its generation, edge computing significantly reduces latency, conserves bandwidth, and improves response times. It ensures that only relevant, processed data is sent to the cloud, optimizing bandwidth and processing power. It complements the cloud computing by offering a balanced approach to data processing. Edge computing addresses latency and bandwidth constraints for real-time applications, whereas cloud computing provides extensive computational power, storage, and advanced analytics.

Examples of smart IoT applications that benefit from edge computing include smart homes, smart vehicles, smart grid, smart cities, smart healthcare, etc. [15]. With regard to the quality attributes, edge computing can contribute to the realization of performance, privacy, energy consumption and bandwidth utilization. Promising features of edge computing include location awareness, mobility support and being close to the user [16]. These features enable edge computing to play a crucial role in addressing different IoT applications such as smart traffic monitoring, industrial automation, virtual reality and smart home. [17].

### 3.2.2. Microservices Architecture

For the last few years, the traditional monolithic approach to the software architecture has fallen out of favour. Its failure to meet the quality requirements of modern systems (e.g. scalability, interoperability and modifiability) has paved the way for the modular architecture approach. This approach has been widely embraced in many domains and become the driver for the development of modern software systems. In this context, Microservices represent a promising architectural model especially in domains where distributed and pervasive computing is highly important, as in Internet of Things [18].

Microservices architecture is built on SOA concepts, focusing on refactoring large monolithic applications into small, highly decoupled, independently deployable and scalable services. These services carry out a single, well defined and domain-oriented task, and communicate among themselves by means of a simple protocol, for instance the HTTP [18] [19].

Quality attributes such as availability, reliability, interoperability, latency, delay, are important characteristics of any distributed system like IoT. Yet, Satisfying and addressing these quality attributes is not a trivial task. The characteristics of microservices make it suitable for addressing this issue at each layer of the IoT architecture [20].

### 3.2.3. Event Driven Architecture

Currently, the development of software systems has taken place in increasingly changing business environments leading to the need for high flexibility to support rapid system adaptations [21]. The real issues of concern in modern software systems development include dynamic software requirements, technology change, a quick response to market, and systems with tightly coupled modules. [22]. Therefore, companies have been searching for better approaches to minimize problems in the development and maintenance of such systems, for instance, the high cost of maintenance and complexity of the systems. Event driven architecture has been largely embraced in the software development, emerging as a promising approach to modular development aiming at supporting the required rapid adaptations of constantly evolving systems [23]. The adoption of event driven architectures aims mainly to improve maintenance by breaking down monolithic applications into independent modules that communicate via events. In the context of IoT systems, the event driven architecture is primarily employed to minimize the coupling between the different components engaged in this ecosystem. It eases the exchange of data between IoT devices while making each one unaware of the others. This facilitates the integration of a new component or device once a change has been required. It also helps in reducing the energy consumption since devices are not constantly sending data to interested parties but instead data are sent only when a significant event is triggered. This is crucial as most IoT devices are resource constrained with regard to energy and capacity. In terms of quality attributes, adopting the event driven architecture helps IoT systems realize the integrability and modifiability. As a messaging protocol, this architectural style uses the publish-subscribe pattern.

## 4. PROPOSED IoT ARCHITECTURE

This section is dedicated to describe the proposed architecture for IoT systems. It includes a meta model that serves as a high-level description of IoT systems. Such a meta model contains the fundamental elements and concepts that form the IoT ecosystem that spread over the different layers of the IoT architecture presented in Figure 1.

### 4.1. A Metamodel for IoT Systems

As a first crucial step towards the design of an effective IoT software architecture, a metamodel is developed. This metamodel defines the main concepts and their relationships and serves as common understanding of the IoT systems. A UML class diagram is employed here to document this metamodel. Following the "divide and conquer" design principle, the metamodel is divided into smaller metamodels, namely the edge computing, the cloud platform and the application.

### 4.1.1. The Edge Computing Metamodel

Figure 2 depicts the concepts that exist in this metamodel as well as the relationships between them.

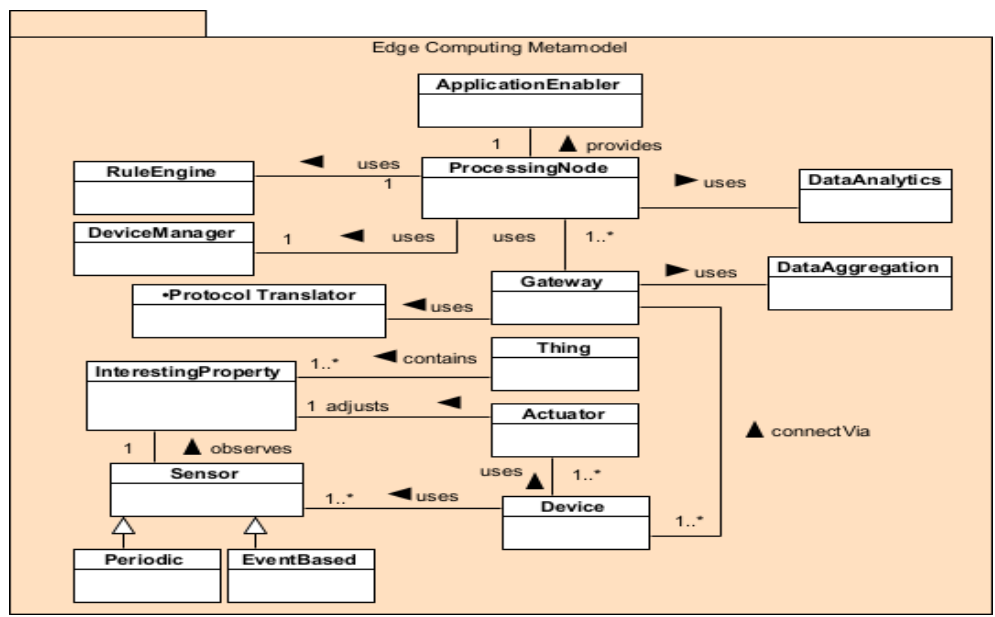

Figure 2. Edge computing metamodel.

These concepts are presented and described as follows:

**Processing Node.** this concept plays the role of the middleware layer of the edge computing. It is responsible for processing data generated and transmitted by a physical device through a gateway. In addition, It also provides a unified APIs to be used by IoT application developers.

**Application Enabler.** this provides the necessary and required APIs for developing IoT applications. Developers can use it to develop

**Rule engine.** This contains a set of rules used by the processing node to control and regulate the operational aspect of IoT applications.

**Data analytics**. It is responsible for analyzing data received from physical devices using some specific analytical tool in order to infer some insights.

**Device manager.** This is responsible for managing IoT devices through neutral and unified interfaces.

**Gateway.** The gateway is the communications hub for the IoT network. It performs important functions, such as data aggregation, translation between communication protocols, and data processing before the gateway forwards the data to the cloud or on-premises network.

**Protocol translator.** Devices send data via non-standard protocols, so they transform it into a common format that the cloud platform can analyze and comprehend with ease.

**Thing.** The "thing" in the IoT vision is very wide concept which includes many forms of physical elements. These include personal objects such as smart phones, tablets and digital cameras as well as elements in our environments (e.g. room, building, vehicle) and industries (e.g., machines, robots). Each thing has attributes that describe it as well as a state that is relevant from a user or an application perspective.

**Interesting property.** It is an observable property of a physical entity. The temperature of a human body is an example of this concept. The property can be single or composed of a number of properties.

**Sensor.** It is a type of resource that has the ability to detect changes in the environment. Thermometer, humidity and motion sensors are some examples of sensor.

**Actuator.** It is a resource that has the ability to make changes in the environment through an action. For example, turn the lights on or the air conditioning system in response to some certain data obtained.

**Device.** It is an entity that hosts sensor and actuator resources and gives them the ability to communicate with other entities. Examples include mobile phones and personal computers.

### 4.1.2. The Cloud Metamodel

As stated earlier in this paper, the edge computing isn't a replacement for cloud computing. Instead, they complement each other and each one has its own use cases and its viable contribution to the delivery of a resilient IoT system. Cloud computing should be used when there is need for heavy data processing and big data need to be stored and analysed using complex data analytics algorithms.

In the proposed architecture, the cloud computing is used as a global controller whose main mission is to regulate the operational aspect of the IoT system in question. This controller accomplishes its task through a set of processing nodes distributed throughout the edge computing platform. The fundamental concepts and elements of cloud metamodel include *global controller, rule engine, data analytics, data storage and processing node*. Such concepts were described in the edge computing metamodel. They have similar tasks except that they operate and coordinate to accomplish the work of the global controller.

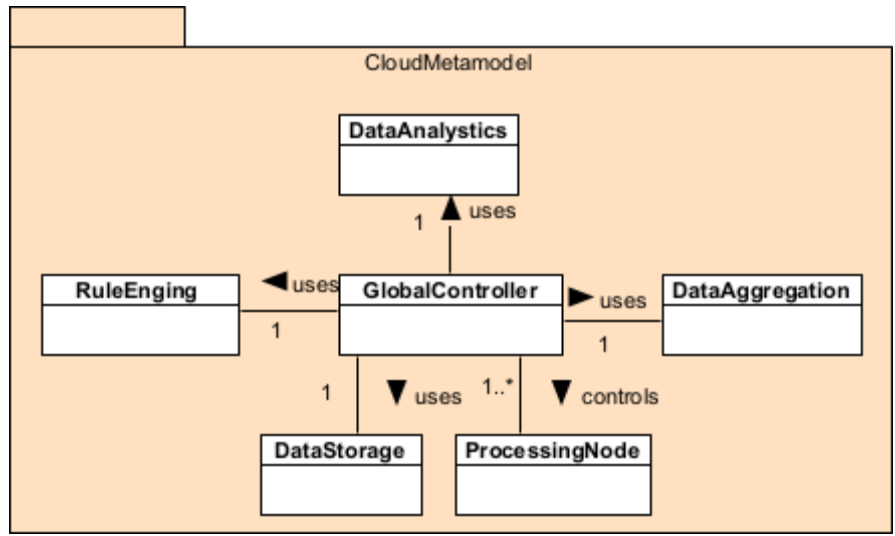

Figure 3. Cloud Metamodel.

### 4.1.3. The Application Metamodel

An IoT application is composed of a set of services to be used by end users. Such services reflect the software requirements of the system and are used to address business values for organizations. Services are provided here to accomplish a number of tasks such as monitoring and controlling devices at the sensing layer, visualizing insights derived from processed data and building an appropriate dashboard to monitor and observe the performance of the whole system. Also, the application user should be brought into the loop by the provision of a user management system that enables the user to create an account which can be used later by the system to accomplish important task such as sending notifications and personalising the user interface of the IoT application. The fundamental concepts and elements of the application metamodel include *device controller, user, user manager, subscriber, dashboard and notifications.*

These concepts are described below.

**Dashboard.** It is the user interface in the IoT platform through which the user monitors and controls connected devices at the sensing layer. It also enables the user to monitor and visualise the application performance and observe any significant changes in the operational aspect either at the device level or the whole system level.

**Notification Manager.** Notifications are sent out to interested users who have subscribed to be notified of some significant events. Therefore, this concept provides the necessary APIs for an interested user to subscribe for a specific event.

**User.** The user is a broad concept that represents the application beneficiary which varies depending on the application domain being developed. An account is created for each user which contains the appropriate information needed for sending out customized notifications and personalising the user interface.

**User Manager.** It is responsible for providing the necessary APIs for creating and managing user accounts.

**Subscriber**. The subscriber is any interested user that registers with the notification manager to receive notifications of some significant events.

**Device controller.** It is used to offer the necessary APIs for controlling a set of available physical devices. Remotely turning on lights at home is an example of using this controller.

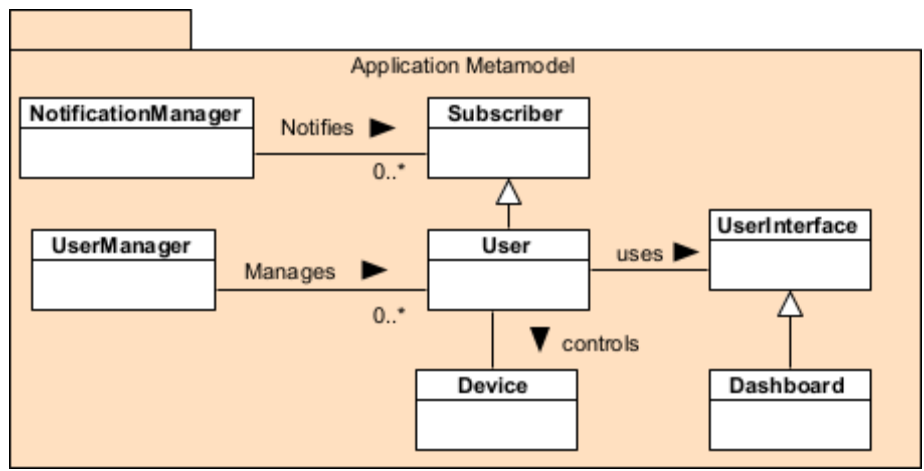

Figure 4. IoT application layer metamodel.

## 4.2. IoT application business logic

In addition to the user interface with which the end user interacts with the IoT system, there is the business logic which constitutes the components of the core system as well as the orchestration system. The latter is responsible for controlling and regulating the operational aspect of the core system.

### 4.2.1. The Core System

In order to model the core system, this paper identifies a number of concepts which are described as follows.

**Domain.** The domain here represents the system in question which comprises a set of tasks. Examples of domain include the healthcare, smart agriculture, home automation, smart metering and smart building.

**Task.** A task is a high-level goal that is addressed in order to realize the overall system requirements. Each task, in turn, encompasses a set of services responsible for achieving that task. A task in a healthcare system is, for example, monitor remotely blood sugar level for a diabetic patient.

**Service.** A service is an abstract representation of a software or hardware entity that plays a role in addressing the ultimate goal of the task at hand. These services are then mapped to software components such as RESTful microservices. A central heating thermostat is an example of service. In our approach, each device or a set of devices involved in IoT applications can be treated as a service.

**Business process.** A Business process represents a specific ordering of services and activities across time and place to realize a high-level business goal. The services of a particular task interact and coordinate with each other to address the purpose of that task. Such coordination is encapsulated in an entity called a business process. A Business process might consist of only one service. However, a useful business process is often composed of a number of services.

### 4.2.2. IoT Orchestration System

To orchestrate the work of the different components of the core IoT system, an appropriate orchestration system must be put in place. The complexity induced by the massive number of connected IoT devices rapidly outperforms the human capability to handle it. As a consequence, the autonomy characteristic is crucial with regard to the design of the orchestration system. Autonomic computing is a field that studies how software systems can achieve desirable behaviours without the human intervention [24]. Such behaviours take the form of self-healing, self-configuring, self-protecting and self-optimising [25].

To design the orchestration system, the IBM autonomic computing model [26] is adopted here. This model uses the MAPE-K feedback control loop which stands for Monitor, Analyse, Plan, Execute and Knowledge. The MAPE-K feedback control loop performs MAPE phases using a shared Knowledge base. The Monitor (M) phase acquires data from the system and its environment through a set of sensors. The Analyse (A) phase involves activities such as filtering and transforming of the monitored data and then analysing the clean data. In the Plan (P) phase, a set of actions are composed taking into account the outcome of the analysis as well as the knowledge of the system. Finally, the Execute (E) of the planned actions should be performed, often through a dedicated number of actuators. A MAPE-K loop stores the Knowledge (K) required for decision making in the Knowledge Base (KB).

The orchestration system is distributed where the local orchestration is performed at the edge computing while the global orchestration is conducted at the cloud platform.

The local orchestration is composed of a set of MAPE-K loops where each loop is in charge of orchestrating a specific region at the edge computing platform.

The local orchestration system is offered as a MAPEaaService in the edge computing platform while the global orchestration is provided as APaaService in the cloud platform since it contains only, in addition to the knowledge component, the analysis and planning activities.

The orchestration system is offered in two modes: centralized and decentralized. In the centralized mode, a central control loop is deployed either on the edge computing or cloud platform (depends on the application scale) to orchestrate the operating of the different control loops that are located at the same level. The master-slave model is employed here to establish such a relationship as shown in Figure 5.

In the decentralized mode, a set of MAPE loops of the same level is coordinated to accomplish the four activities (monitoring, analysis, planning and execution). For instance, the execute components of involved control loops communicate and coordinate to perform the corrective actions in the absence of a central controller. Self-organising systems are a popular example of systems operating and functioning in the decentralised mode.

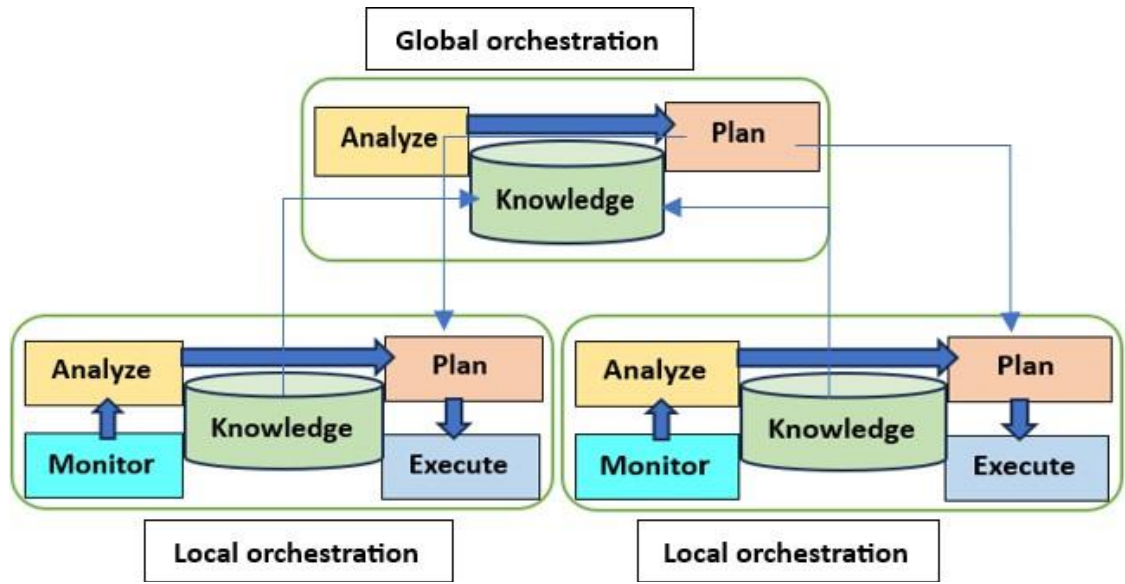

Figure 5. The centralized mode of the orchestration system.

## 4.3. Proposed Architectural and Design Patterns

In this section, two important architectural and design patterns are presented to demonstrate the realization of some of the important quality attributes highlighted earlier in this paper. A description of these patterns is presented as follows.

- **The Master Slave pattern**: it is a design model in computing where one central entity, called the master, controls and directs the operation of multiple subordinate entities, known as slaves. In the proposed architecture, this pattern was employed to establish the architectural relation between the global orchestration system and a set of local orchestration systems. This architectural pattern enables parallel processing and load balancing, thereby improving system performance and scalability.

- **The Publisher-Subscriber pattern:** it is a messaging pattern where publishers send messages without knowing who will receive them, and subscribers receive messages without knowing who sent them. The communication between publishers and subscribers is facilitated by a third party known as message broker. This decoupling allows for more flexible and scalable systems, as components can evolve independently.

  In the proposed architecture, this pattern may be used to model the notification system where interested users subscribe to interesting events categorized as topics in the broker. The publisher here can be any event source such as a senor of a specific device or a monitor in the MAPE-K loop that publishes a new data to the knowledge base after which the analyse component gets notified. These two scenarios are depicted in Figures 6 and 7.

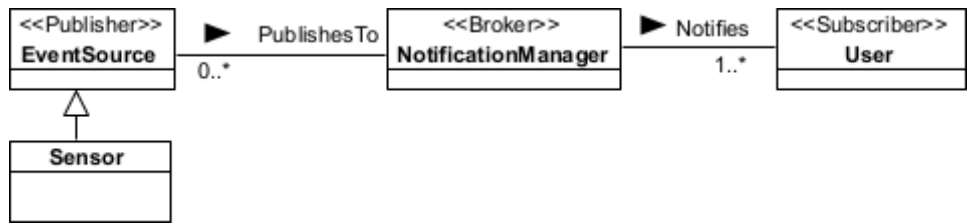

Figure 6. Publisher-Subscriber patten at the IoT application layer.

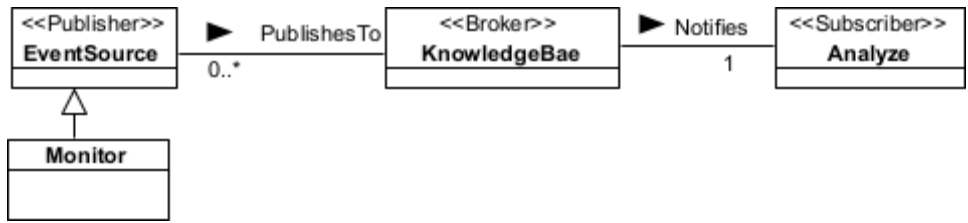

Figure 7. Publisher-Subscriber patten at the local orchestration System.

The Publisher-Subscriber model is a powerful pattern that is adopted in modern software development to address many challenges. By decoupling components, it enables scalable, maintainable, and flexible systems suitable for IoT applications.

## 5. RELATED WORK

This section presents some works related to the research conducted in this paper. Therefore, works on software architecture, architectural patterns and design patterns for IoT systems are considered.

Stojanov et al. [11] presented discussion of specific quality attributes during the design of layered sensor-based IoT system for monitoring the industrial environmental conditions in manufacturing sector. They emphasized on the importance of considering quality attributes of software architecture in order to ensure proper functioning of the whole systems and minimizing risks and costs in the system life cycle. Kim [28] proposed a quality model of IoT applications based on the following characteristics of IoT systems: participation of hardware devices, collaboration of software and hardware components, mobility and connectivity of devices, monitoring of devices, and limited resources. Temkar et al. [29] conducted a study to evaluate the quality of a system for agriculture field monitoring using Wireless Sensor Network and IoT. The system architecture consists of components that enable data collection and monitoring, data processing, execution, and feedback.

Guth et al. [7] proposed an abstract IoT reference architecture in an attempt to enable a uniform terminology and to ease the comparison of existing platforms. Porruvecchio et al. [27] introduced an IoT platform called CMC-IoT. Such a platform is based on the microservices architecture aiming at providing tools for developing scalable and robust IoT applications.

In [30], The IoT-A project proposed an architectural reference model and a preliminary set of buildings blocks to promote a fully interoperable and scalable vision of IoT. The foundation of the IoT-A Reference Model is the IoT Domain Model, which introduces the main concepts of the Internet of Things. Important concepts include Devices, IoT Services, and Virtual Entities (VE), as well as the relations between these concepts. Qanbari et al. [31] proposed design patterns for edge computing based applications. They include: 1) edge computing pattern to handle the provision of all edge devices automatically; 2) source code deployment pattern for edge devices to handle the deployment of the code to all devices connected to the IoT system; 3) edge orchestration pattern to handle the automation of creation, monitoring, and deployment of resources in the IoT environment.

Despite the significant contribution made by the papers presented above, most of them overlooked the importance of building software architectures driven by quality attributes where the focus is on the rationale choice of software components and the way they relate to each other.

# 6. CONCLUSIONS

This paper was an attempt to lay out the foundation for a generic software architecture for IoT platform. The work presented here emphasizes the importance of a well designed software architecture driven by some related quality attributes for the delivery of a resilient and future proof IoT systems. Significant architectural styles and design patterns were proposed to address these quality attributes. These include edge computing, microservices, event driven architectures, and the publish-subscribe pattern (based on the observer design pattern). The emphasis was also on treating IoT systems as autonomic systems which require a closed control loop to regulate and orchestrate the operational aspect of the IoT system. The MAPE-K control loop proposed by the IBM autonomic computing model was adopted in this paper. Future works may include a more detailed and technical description of each concept and component involved in the metamodel of the proposed architecture. Also, a dedicated case study should be conducted to clarify and demonstrate the application of proposed architectural and design patterns.

## AUTHOR

**Yousef Abusetta** is an associative professor at the University of Tripoli, Libya. He joined the Internet Technologies department of the IT faculty in 2022. He is currently in charge of the supervision of the postgraduate studies program at the Internet Technologies department. His research interests include autonomic systems, IoT, software architecture and design patterns.